\newcommand{\be}{\begin{equation}}
\newcommand{\ee}{\end{equation}}
\newcommand{\bea}{\begin{eqnarray}}
\newcommand{\eea}{\end{eqnarray}}
\newcommand{\ben}{\begin{eqnarray*}}
\newcommand{\een}{\end{eqnarray*}}
\documentclass{elsart5p}

\usepackage{graphicx,color}
\usepackage{dcolumn}
\usepackage{bm}
\usepackage{amsmath}
\usepackage{amssymb}
\usepackage{mathrsfs}
\usepackage{subfigure}
\begin{document}
\begin{frontmatter}
\title{Entanglement in the 1D extended anisotropic Heisenberg model}
\author{E. Plekhanov\corauthref{Plekhanov}},
\ead{plekhanoff@physics.unisa.it}
\author{A. Avella},
\author{F. Mancini}
\address{Dipartimento di Fisica ``E.R. Caianiello'' - Unit\`{a}
CNISM di Salerno, Universit\`{a}
degli Studi di Salerno, I-84081 Baronissi (SA), Italy}
\corauth[Plekhanov]{Corresponding author. Tel: +39 089 965228
fax: +39 089 965275}
\begin{abstract}
We present a study of entanglement in the case of the 1D extended
anisotropic Heisenberg model. We investigate two quantum phase
transitions (QPTs) within the previously found ergodicity phase diagram [E.
Plekhanov, A. Avella, and F. Mancini Phys. Rev. B \textbf{74}, 115120
(2006)]. Our calculations are done by means of the numerically exact
Lanczos method at $T=0$, followed by a finite-size scaling. As a
measure of entanglement we use the concurrence for two spins out
of the system. We conclude from our studies that these QPTs are
accompanied by a qualitative entanglement change.
\end{abstract}
\end{frontmatter}
\section{Introduction}
Quantum phase transitions (QPTs) are among the most fascinating topics
in quantum mechanics. Recently, in connection to the quantum
information theory (QIT), the correlation between QPTs and
entanglement has been extensively studied~\cite{review}. Many systems
have been found where QPT is accompanied by a qualitative change
of the entanglement, suggesting an implicit connection between
the two.
The spin systems are a natural implementation (at least
theoretically) for QIT devices due to
the isomorphism between the Hilbert spaces of a single spin and the
one of a
qubit - the central object of quantum information. From the QIT point
of view, the spin chains represent a non-trivial example of the qubit
network, and it is therefore legitimate to look for the entangled
states of such systems. There exist a few entanglement measures, which
differ mainly in the way the system is split into blocks,
whose entanglement is measured. The one-tangle, or von Neumann
entropy, is only a function of the local magnetisation and, hence, is
scarcely informative. It is therefore more preferable
to use the concurrence~\cite{wootters}, or pairwise entanglement, which, on the
contrary, depends on the spin-spin correlation functions.
The concurrence for a couple of spins at sites $i$ and $j$ is defined as:
\be
   C_{i,j} = \textrm{max}( 0, \lambda_1 - \lambda_2 - \lambda_3 -
   \lambda_4 ),
   \label{def_cunc}
\ee
where $\{\lambda_i\}$ are the eigenvalues, in decreasing order, of the
Hermitian matrix $R$, which can be expressed in terms of the
two-point correlation functions~\cite{wootters}. Its extreme values, zero
or one, indicate that the system is either a product state or
maximally entangled, respectively. In order to take
into account the contribution to the entanglement, coming from the
correlations at all distances we adopt $\tau_2$~\cite{coffman} as a
measure of the strength of the entanglement:
\be
   \tau_2 = \sqrt{
   \sum_{d > 0} C^2_{i,i+d},
   }
\ee
where $C_{i,i+d}$ is the concurrence between the site $i$ and its
$d$-th neighbor. In a translationally invariant state $\tau_2$ does
not depend on $i$.
\begin{figure*}
\begin{center}
\mbox{
\subfigure[{Phase diagram of (\ref{ham}) in the $J^{\prime}-J_{\perp}$ plane.}\label{fig:phd}]
{\includegraphics[angle=0,width=0.33\textwidth]{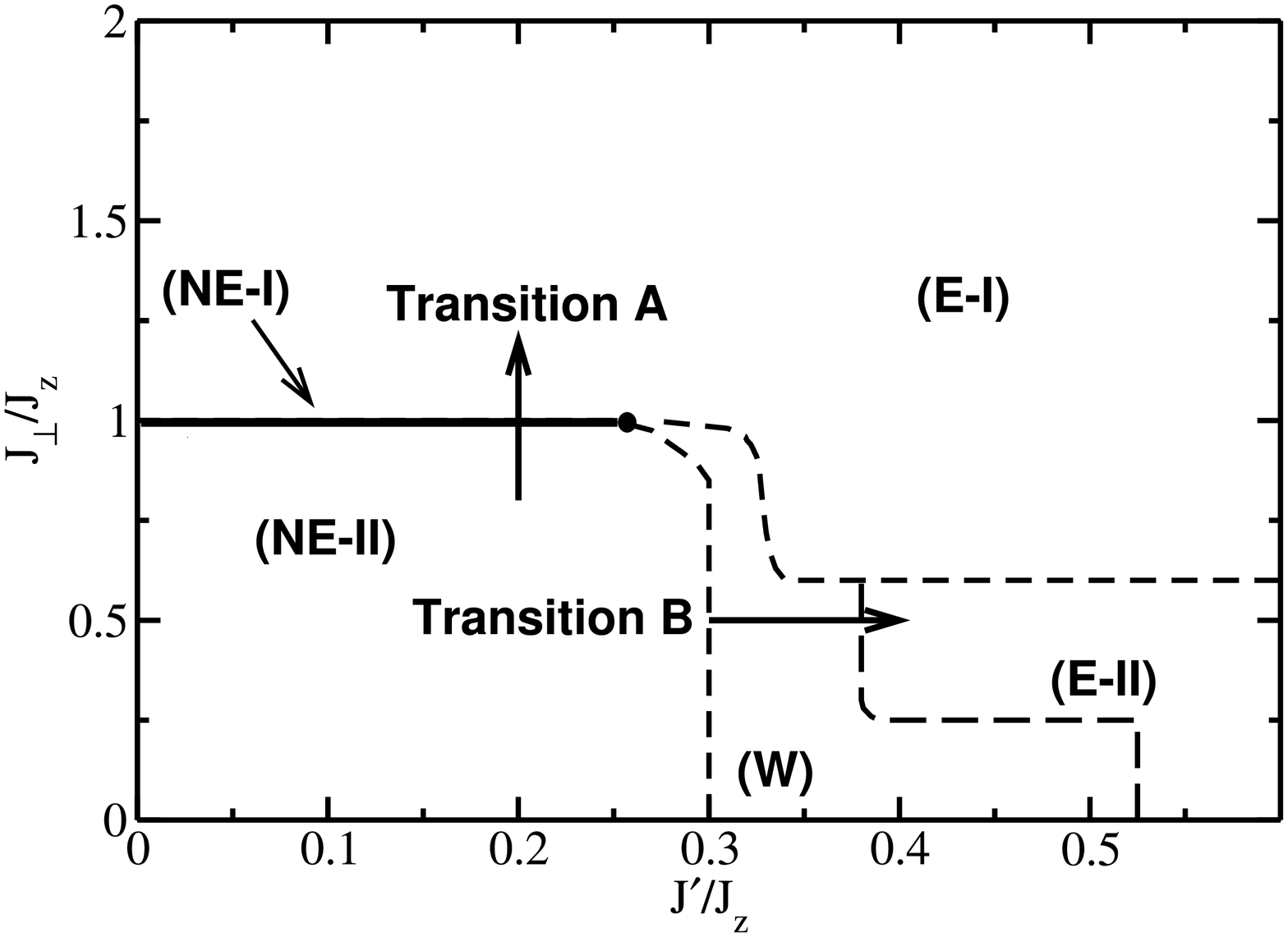}}\hspace{0.25cm}%
\subfigure[{Transition A: $J^\prime$ is set to $0.2J_z$}\label{fig:tra}]
{\includegraphics[angle=0,width=0.34\textwidth]{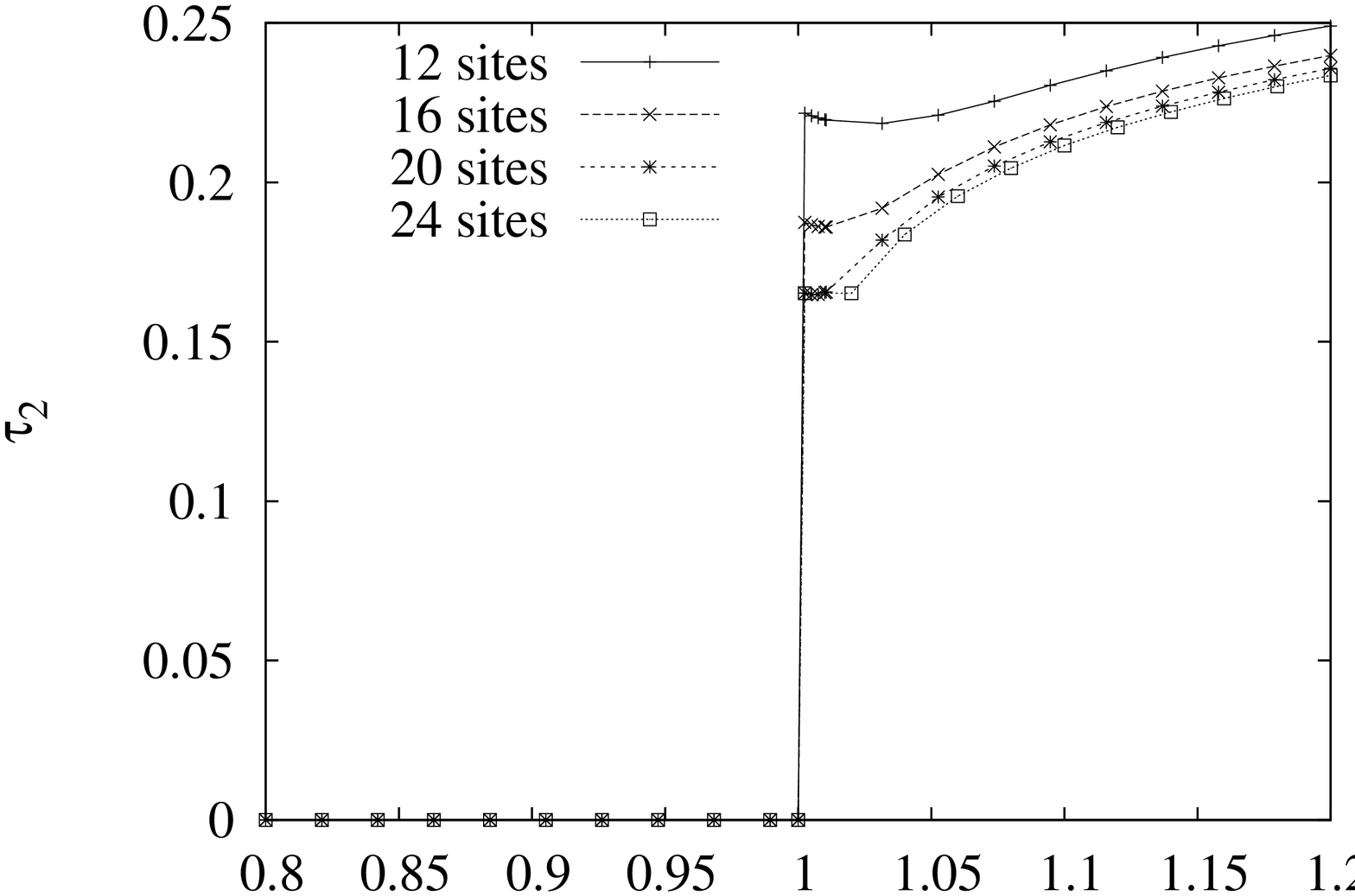}}
\subfigure[{Transition B: $J_{\perp}$ is set to $0.5 J_z$}\label{fig:trb}]
{\includegraphics[angle=0,width=0.34\textwidth]{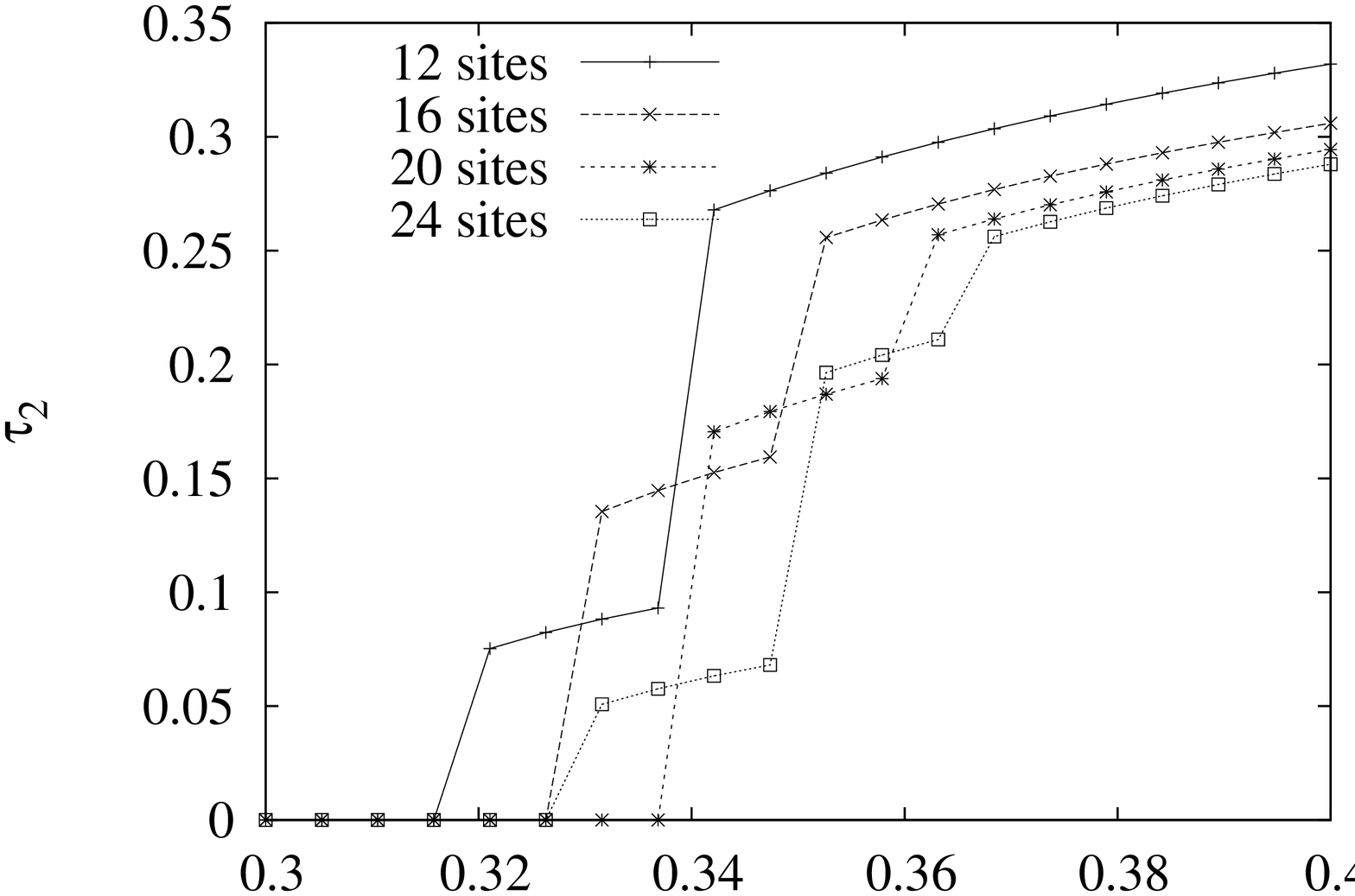}}
}
\caption{
(a) Schematic phase diagram of the model (\ref{ham}). See Ref.~\cite{ours} for details.
Notice the locations and directions of the lines crossing the two phase transitions.
Entanglement measure $\tau_2$ across transition A (a) and transition B (b)
for different system sizes $L=12,16,20,24$. 
}
\end{center}
\end{figure*}
\section{Model and Method}
We measure $\tau_2$ in the 
one-dimensional anisotropic extended Heisenberg model
with next-nearest-neighbor interaction on a chain with 
$L$ sites, subject to periodic boundary conditions:
\bea
   \nonumber
   H &=& -J_z \sum_i S^z_i S^z_{i+1} + J_{\bot} 
   \sum_i ( S^x_i S^x_{i+1} + S^y_i S^y_{i+1}) \\
     &+&  J^{\prime}\sum_i \mathbf{S}_i \mathbf{S}_{i+2}.
   \label{ham}
\eea
We diagonalize the Hamiltonian (\ref{ham}) by means of the
Lanczos diagonalization technique.  In doing that we take into account
translational symmetry and classify the eigenstates by the eigenvalues of
total $S^z$ which is a good quantum number. We use $J_z>0$,
which corresponds to ferromagnetic coupling.
\section{Results}
The phase diagram in the $J^{\prime}-J_{\perp}$ plane, as found in
Ref.~\cite{ours} and shown
on Fig.~\ref{fig:phd}, manifests the presence of a totally polarized phase
(NE-II phase), which neighbors two paramagnetic phases at the lines
$J_{\perp}=J_z$ and $J^{\prime}\approx 0.33 J_z$. In this article we
study these two phase transitions, which we will call transition A and
transition B, respectively. In the NE-II the ground state is doubly
degenerate with all spins either "up" or "down" and therefore
$\tau_2=0$ because this is a product state. The behavior
of $\tau_2$ across the transition lines for two representative values
of $J^\prime$ and $J_\perp$, is shown in Fig.~1(b)-(c), respectively.
We summarize the properties of the two transitions as follows:

\textit{Transition A}:
In this transition frustration, induced by the antiferromagnetic
interaction, parametrized by $J_\perp$, destroys the ferromagnetic
order. If $J_\perp>J_z$ the system becomes antiferromagnetic in $XY$
plane, as emerges from our analysis~\cite{new_article}. It is clear
that such a ground state is entangled. $C_{i,i+d}$ monotonically
decreases as a function of $d$, and the main contribution to $\tau_2$
comes from the nearest-neighbor term ($d=1$).  For the system sizes
considered, $L=12\div 24$, the decay of the antiferromagnetic
correlations is very weak and they scale well with increasing $L$.
This fact explains the good finite-size scaling of our results for
$\tau_2$.

\textit{ Transition B}:
In the definition of the concurrence~(\ref{def_cunc}) it might happen
that for some $d$ there is no dominant eigenvalue of $R$ and
$C_{i,i+d}=0$.  This would mean that for this particular $d$ the two
spins are untangled from the rest of the system. This is the case of the
phase, situated behind the transition B. Precisely, the only non-zero
contribution to $\tau_2$ comes from $d=2$. This implies the enhancement
of correlations at $d=2$, as indeed found by examining the correlation
functions.  This feature is short-ranged even for the small sizes
considered here and persists both in $Z-$direction and in $XY$ plane.
Within the range of $J^{\prime}$ considered [$0.3J_z,0.4J_z$] there are
regions, characterized by a finite magnetisation per site less than
$1/2$. For $J^{\prime}\gtrsim 0.37J_z$ the magnetisation stabilizes at
zero. It is this finite magnetisation which is responsible for the
steps of $\tau_2$ on the right panel of Fig.~\ref{fig:trb}.
Unfortunately our computation facilities did not allow us to conclude
whether these steps are a finite-size effect, and whether the vanishing
of the ground state magnetisation at the QPT is continuous or not. The
shorter range of the correlations respect to the Transition A case and
the presence of the magnetisation steps imply a worse finite-size
scaling in this case.

 In conclusion, we have studied the entanglement change across two QPTs
in the vicinity of the ferromagnetic phase 
in the extended anisotropic Heisenberg model~(\ref{ham}).
The entanglement appears to be intimately connected to the underlying
ground state of the system. Starting from a totally untangled
ferromagnetic state, by changing the Hamiltonian parameters, we observed
the complete reordering of the ground state structure. Such reordering
is accompanied by the appearance of pairwise entanglement, reflecting the
structure of underlying ground state. Namely, in the transition A,
entanglement is non-zero for all the distances, being maximal for
nearest neighbors, while for transition B only the next-nearest-neighbor
entanglement is non-zero.

\end{document}